\documentstyle[psfig]{mn}

\newif\ifAMStwofonts

\def\h0{{\rm H_0}}
\def\mum{\mu {\rm\,m}}
\def\pm{^+_-}

\ifoldfss
  \ifCUPmtlplainloaded \else
    \NewTextAlphabet{textbfit} {cmbxti10} {}
    \NewTextAlphabet{textbfss} {cmssbx10} {}
    \NewMathAlphabet{mathbfit} {cmbxti10} {} 
    \NewMathAlphabet{mathbfss} {cmssbx10} {} 
  \fi
  \ifAMStwofonts
    \ifCUPmtlplainloaded \else
      \NewSymbolFont{upmath} {eurm10}
      \NewSymbolFont{AMSa} {msam10}
      \NewMathSymbol{\upi}     {0}{upmath}{19}
      \NewMathSymbol{\umu}     {0}{upmath}{16}
      \NewMathSymbol{\upartial}{0}{upmath}{40}
      \NewMathSymbol{\leqslant}{3}{AMSa}{36}
      \NewMathSymbol{\geqslant}{3}{AMSa}{3E}

    \fi
  \fi
\fi 

\ifnfssone
  \newmathalphabet{\mathit}
  \addtoversion{normal}{\mathit}{cmr}{m}{it}
  \addtoversion{bold}{\mathit}{cmr}{bx}{it}
  \newmathalphabet{\mathbfit} 
  \addtoversion{normal}{\mathbfit}{cmr}{bx}{it}
  \addtoversion{bold}{\mathbfit}{cmr}{bx}{it}
  \newmathalphabet{\mathbfss} 
  \addtoversion{normal}{\mathbfss}{cmss}{bx}{n}
  \addtoversion{bold}{\mathbfss}{cmss}{bx}{n}
  \ifAMStwofonts
    \ifCUPmtlplainloaded \else
      \UseAMStwoboldmath
      \makeatletter
      \new@mathgroup\upmath@group
      \define@mathgroup\mv@normal\upmath@group{eur}{m}{n}
      \define@mathgroup\mv@bold\upmath@group{eur}{b}{n}
      \edef\UPM{\hexnumber\upmath@group}
      \new@mathgroup\amsa@group
      \define@mathgroup\mv@normal\amsa@group{msa}{m}{n}
      \define@mathgroup\mv@bold\amsa@group{msa}{m}{n}
      \edef\AMSa{\hexnumber\amsa@group}
      \makeatother
      \mathchardef\upi="0\UPM19
      \mathchardef\umu="0\UPM16
      \mathchardef\upartial="0\UPM40
      \mathchardef\leqslant="3\AMSa36
      \mathchardef\geqslant="3\AMSa3E
    \fi
  \fi
\fi 

\ifnfsstwo
  \DeclareMathAlphabet{\mathbfit}{OT1}{cmr}{bx}{it}
  \SetMathAlphabet\mathbfit{bold}{OT1}{cmr}{bx}{it}
  \DeclareMathAlphabet{\mathbfss}{OT1}{cmss}{bx}{n}
  \SetMathAlphabet\mathbfss{bold}{OT1}{cmss}{bx}{n}
  \ifAMStwofonts
    \ifCUPmtlplainloaded \else
      \DeclareSymbolFont{UPM}{U}{eur}{m}{n}
      \SetSymbolFont{UPM}{bold}{U}{eur}{b}{n}
      \DeclareSymbolFont{AMSa}{U}{msa}{m}{n}
      \DeclareMathSymbol{\upi}{0}{UPM}{"19}
      \DeclareMathSymbol{\umu}{0}{UPM}{"16}
      \DeclareMathSymbol{\upartial}{0}{UPM}{"40}
      \DeclareMathSymbol{\leqslant}{3}{AMSa}{"36}
      \DeclareMathSymbol{\geqslant}{3}{AMSa}{"3E}
    \fi
  \fi
\fi 

\ifCUPmtlplainloaded \else
  \ifAMStwofonts \else 
    \def\upi{\pi}
    \def\umu{\mu}
    \def\upartial{\partial}
  \fi
\fi

\title{Through a Lens Darkly: Evidence for Dusty Gravitational Lenses}

\author[Malhotra, Rhoads, \& Turner]
{Sangeeta Malhotra,$^1$
James E. Rhoads,$^2$
Edwin L. Turner,$^3$\\
$^1$IPAC, M.C. 100-22, California Institute of Technology, Pasadena, CA 91125, USA.\\
$^2$Kitt Peak National Observatory, 950 N. Cherry Avenue, Tucson, AZ 85726, USA. \\
$^3$Princeton University Observatory, Princeton NJ 08540, USA.}

\date{Accepted 1996
      Received 1996 October 31
      in original form 1996 October 31}

\pagerange{\pageref{firstpage}--\pageref{lastpage}}
\pubyear{1996}

\begin{document}

\maketitle

\label{firstpage}

\begin{abstract}
Foreground galaxies that amplify the light from background quasars may
also dim that light if the galaxies contain enough dust. Extinction by
dust in lenses could hide the large number of lensed systems predicted
for a flat universe with a large value of the cosmological constant
$\Lambda$.  We look for one signature of dust, namely reddening, by
examining optical-infrared colors of gravitationally lensed images of
quasars. We find that the lensed systems identified in radio and
infrared searches have redder optical-IR colors than optically
selected ones.  This could be due to a bias against selecting reddened
(hence extincted) quasars in the optical surveys, or due to the
differences in the intrinsic colors of optical and radio quasars.
Comparison of the radio-selected lensed and unlensed quasars shows
that the lensed ones have redder colors.  We therefore conclude that
at least part of the color difference between the two lens samples is
due to dust.

From the color difference between lensed and unlensed radio quasars
(and assuming Galactic extinction law) we can reconcile a large
cosmological constant ($\Lambda=0.9$) with the number of lensed
systems observed in flux limited optical surveys. These results
substantially weaken the strongest constraint on cosmological
scenarios that invoke a non-zero cosmological constant to explain age
discrepancy problems, satisfy predictions of inflationary models of
the early universe and play a role in large scale structure formation
models.  They also raise the prospect of using gravitational lenses to
study the interstellar medium in high redshift galaxies.
\end{abstract}


\section{Introduction}

The statistics of gravitational lensing provide some of the strongest upper
limits on the cosmological constant $\Lambda$ (Turner 1990, Fukugita,
Futamase, \& Kasai 1990), one of the three parameters ($\Omega, \Lambda, \h0$)
that determine the geometry and the age of the universe (Carroll, Press \&
Turner 1992). Models with large values of $\Lambda$ predict many more lensed
objects than are presently observed in optical surveys (Maoz \& Rix 1993,
Kochanek 1993). The constraints from radio surveys and radio lenses are not so
stringent because the redshifts and the luminosity function of radio sources
are not well determined (Kochanek 1996).

The idea that the dust in lensing galaxies can be an important factor is
fairly recent. Fukugita \& Peebles 1993 calculated the effect of dust on
lensing statistics by making all ellipticals dusty beyond a redshift of
0.5. Kochanek (1996) showed that one would need an average reddening of 2
magnitudes to reconcile the lensing statistics with a high $\Lambda$
universe. Lawrence et al. (1995) and Larkin et al. (1994) show examples of
heavily reddened lensed quasars found with the MIT-Green Bank survey and
Nadeau et al. 1991 demonstrated differential reddening between the images of
Q2237+0305, which is lensed by a spiral.  The idea of dust in lensing galaxies
may have been slow in coming because most of the lensing at arcseconds scales
is due to elliptical and S0 galaxies (Turner Ostriker \& Gott 1984 (TOG84))
believed to have little or no interstellar matter at the present epoch. There
is however evidence for some dust and gas in present day early type
galaxies. More than 50\% of elliptical and S0 galaxies are detected in dust
emission with IRAS (Knapp et al. 1989) and $\simeq 40 \%$ of elliptical
galaxies from the Revised Shapely-Ames catalog have dust lanes or patches at
optical depths $0.1 < \tau_V < 0.7$ (Goudfrooij \& de Jong 1995). Dust lanes
were seen in the cores of 78 $\pm$ 16\% of elliptical galaxies observed with
HST (van Dokkum \& Franx 1995).

Besides misleading us about the total number of gravitational lenses, dust in
lenses may bias our conclusions regarding the relative number of lenses as a
function of image separation, another diagnostic of cosmology and structure
formation models (TOG84, Wambsganss et al. 1995). Also, higher redshift lensing
galaxies will have more extinction for the same amount of dust because they see
the quasar light at shorter wavelengths. This may bias the tests for $\Lambda$
based on the redshifts of the lensing galaxies (Kochanek 1992).

In section 2 we report a purely empirical test to determine whether there is
dust in lenses by comparing the optical-IR colors of radio-selected lens
systems with those of optically selected systems. Dust transmits redder
wavelengths preferentially, so objects seen through large amounts of dust are
reddened and dimmed in optical wavelengths. The faint, reddened images may
easily be missed by optical surveys.  At radio wavelengths, dust is
transparent, and we expect radio surveys to contain lens systems with all
degrees of reddening, without bias.  The radio selected sample should therefore
be redder if there is a significant amount of dust in the lensing galaxies.

In presenting evidence for reddening of the lensed image due to the
intervening lensing galaxy, we need to address two issues: (1) Are the lensed
radio-selected quasars redder than the optically selected lensed quasars,
since we expect not to be able to observe many reddened optical lenses? (2)
Are these lenses redder than the unlensed population they are derived from?
These two questions are related; the optical sample from which the lenses are
identified may be intrinsically less red than the radio sample, and that
could well explain the color difference between the two lensed samples.  In
section 3 we address the second question by comparing the optical-IR colors
of lensed and unlensed radio and optical quasars.

In section 4 we discuss other effects besides dust reddening which can result
in red optical-IR colors; some other evidence for dust and interstellar medium
in lensing galaxies; implications for using lensing optical depth as a test for
$\Lambda$; and further tests to verify the presence of dust in lensing
galaxies.

\section{Optical-Infrared colors of Lensed Quasars}

We compare the optical-infrared colors of the radio sample with the optical
sample.  The infrared and optical photometry of an almost complete sample of
known and suspected cases of multiply imaged objects (Keeton \& Kochanek
1996, Surdej 1996) is compiled partly from an infrared survey of
gravitational lenses carried out at Apache Point Observatory (Rhoads et
al. 1997) and partly from the literature (Soifer et al., Lawrence et
al. 1995, Annis \& Luppino 1993, Larkin et al. 1994, Nadeau et al. 1991,
Lawrence et al. 1992, 1995, Barvainis et al. 1995, Rowan-Robinson et al 1993,
McLeod, Rieke \& Weedman 1994, Tinney 1995).

Figure 1 shows the distribution of the optical-IR2 colors and the IR1-IR2
colors of the two samples.  The optical band is mostly R and the two infrared
bands are J(IR1) and H or K (IR2).  The wavelengths observed are slightly
inhomogeneous but that is negligible compared to the wide range of emission
wavelengths due to the broad redshift distribution of the sources.  We also
include in the sample a single lens, FS10214+4724, found in the IRAS survey
(and formerly the most luminous object in the universe) and treat it as a radio
selected source since it was not subject to the biases that the optical surveys
may have been.

\begin{figure}
\psfig{figure=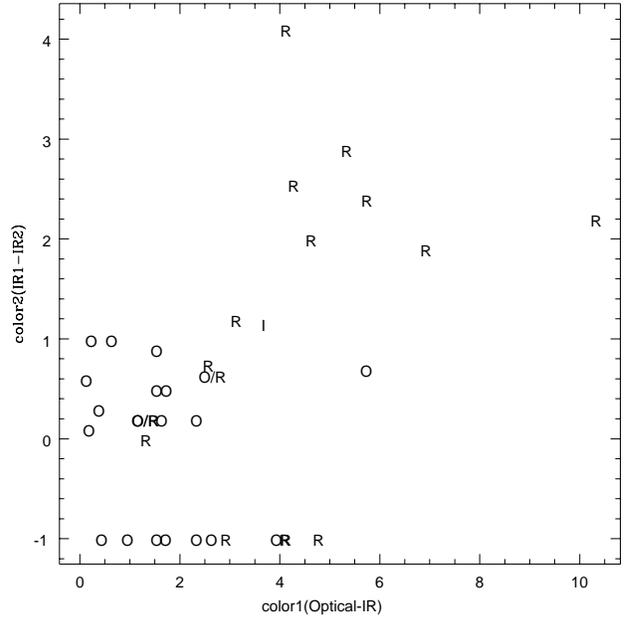,width=3.5in}
\caption{The optical-infrared and infrared-infrared colors of known and
 suspected multiply imaged gravitationally lensed objects. The optical band is
 mostly R and sometimes r or i. The near infrared bands are J for IR1 and H or
 K for IR2. The radio selected objects are represented by R and optically
 selected objects are denoted by O. I denotes the sole infrared selected object
 IRAS F10214+4724. We see that compared to the radio selected sources optically
 selected lenses are systematically bluer in both colors. In fact the color
 distribution of optical and radio samples are almost disjoint. The trends are
 consistent in the two colors shown, i.e. objects red in Optical-IR2 color are
 red in IR1-IR2 color as well. Where we have a measurement of only one infrared
 band we denote the IR1-IR2 color as -1.}
\end{figure}

It is clear from Figure 1 that the radio selected lensed images are redder than
the optically selected ones. The optically selected and radio-selected quasars
have almost disjoint distributions in optical-IR colors. The Wilcoxon test
(Lupton 1993) for the means of the two distributions shows that the mean
optical-IR color of the radio selected sample is different from the color of
the optical sample at 99.99\% confidence level. The IR1-IR2 colors are likewise
very different in the different samples.

We have used each resolved image or unresolved group of images as an
independent point, which is justifiable if the different color is due to
reddening and each image goes through a different path in the lensing galaxy.
If the color difference between the radio and optical samples is due to the
intrinsic color difference in the background sources, then each source, and
not each image, is an independent data point. In that case the mean
optical-IR colors of the two samples differ at the 99.7\% confidence level.

Since the different lensed objects lie at different redshifts and are measured
in a slightly inhomogeneous set of bands, the best way to compare their colors
is by means of the spectral index $\alpha$, where $f_{\nu} \propto
\nu^{\alpha}$ and $f_{\nu}$ is the flux density at frequency $\nu$. (This is a
useful quantity because quasar spectra are power laws to a good approximation
at these wavelengths). Since most of the sources are at redshifts between 1 and
4, we are actually measuring ultraviolet-optical colors in the source rest
frame.  Figure 2 shows the spectral indices for the radio and the optical
samples.  We also calculate the spectral indices that would be expected for
each lensed object if it had the mean spectral energy distribution of an
unlensed radio-quiet or radio-loud quasar (Elvis et al. 1994).

\begin{figure}
\psfig{figure=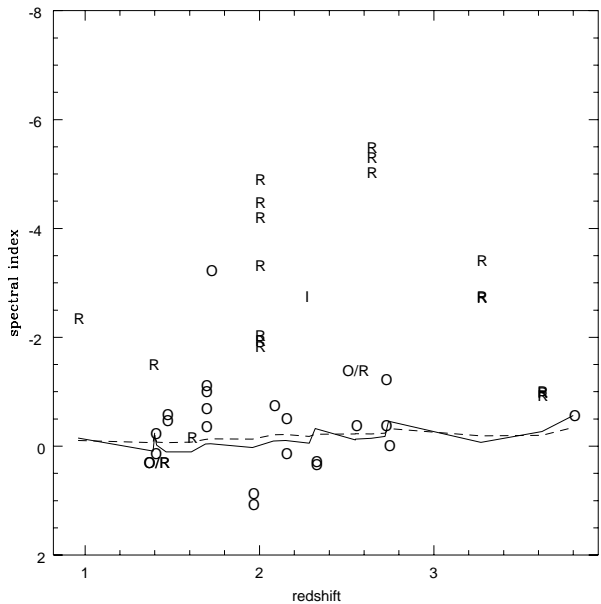,width=3.4in}
\caption{The spectral index of the lensed quasars is plotted
 against their redshift. The radio selected objects are represented by R and
optically selected objects are denoted by O. I denotes the sole infrared
selected object IRAS F10214+4724. The solid line shows the spectral index for
the lensed object if it had the mean SED (Elvis et al. 1994) of an unlensed
radio-loud quasar. The dotted line shows the same quantity if the source was
a radio-quiet quasar. The optically selected lensed quasars have spectral
indices clustered around the mean index of unlensed quasars, and the radio
selected sample shows steeper spectra. For some sources the redshifts are not
known; they are plotted here at a redshift of 2.}
\end{figure}

We see that the radio sample has steeper spectral indices than the optical
sample and that the redshift distribution of the two samples is very similar
(Wilcoxon test shows that the mean redshifts differ only at the 6\%
confidence level). So it is unlikely that any special spectral features
contribute to the color difference of the radio and the optical samples. The
spectral indices derived from the optical lensed sample in general match well
with those derived from the mean Spectral Energy Distribution (SED) from both
the radio loud and the radio quiet samples, whereas the spectral indices of
the radio lensed sample indicate redder colors.

\section{Comparison with unlensed quasars}

The lens systems come from many surveys and other observations, notably optical
surveys PG (Green, Schmidt \& Liebert 1986), LBQS (Hewett et al 1995), and
radio surveys MG (Burke et al. 1992), JVAS (Patnaik 1994, King et al. 1996),
CLASS (Myers 1996). Because lensed sources are magnified we cannot present an
exhaustive survey of the optical-IR colors or optical spectral indices of these
surveys down to the intrinsic faintness of the lensed objects. We can however
compare the spectral indices (and hence the colors) of the lensed samples with
some unlensed objects in the samples (Sanders et al. 1989, Francis, 1995).
None of the radio surveys have systematic measurements in optical and IR. The
JVAS survey consists of flat spectrum radio sources and the MG survey does not
discriminate on the basis of spectral index.  We use two samples of unlensed
quasars with flat and steep radio spectra (Webster et al. 1996, Dunlop et al
1989) . Comparing the spectral indices derived from these two samples
separately by Wilcoxon test, we conclude that the radio selected lensed quasars
have steeper spectral indices than the radio selected quasars at the 99.99\%
confidence level. The optically selected lensed quasars show no significant
difference in their spectral indices as compared to the unlensed sample.

Figure 3 shows the distribution of the spectral indices for the 6 samples: The
Palomar Green (PG) survey, Large Bright Quasar Survey (LBQS), optically
selected lensed quasars, the Parkes radio quasars (Webster et al. 1995, Dunlop
et al. 1995) and the radio selected lensed objects. The spectral indices were
computed (mostly) from $R (0.7 \mum)$ and $K (2.2 \mum)$ broadband
photometry. The optically selected lenses and PG quasars are not very different
in their spectral indices. The radio selected lensed sample differs
significantly from the unlensed Parkes survey sample, with the spectral index
being steeper (implying redder colors) for the lensed quasars.  The (unlensed)
radio quasars are redder than the optical PG and LBQS quasars. Webster et
al. (1995) interpret the red color of some of the Parkes quasars as evidence of
dust reddening between the quasar and the observer. This is consistent with
reddening being more frequent for lensed objects, since their light has {\it a
posteriori\/} a much higher chance of encountering a fully formed galaxy with
dust in it.

\begin{figure}
\psfig{figure=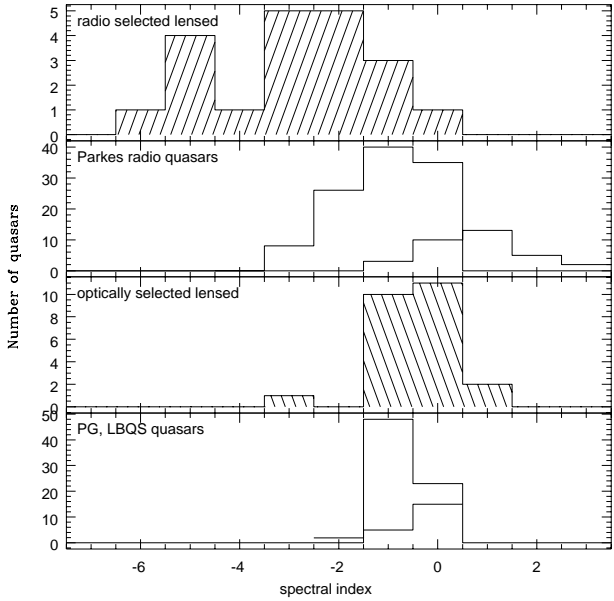,width=3.4in}
\caption{ A histogram of the distribution of spectral indices in various
samples of quasars. Going from the bottom panel to the top we plot optical
unlensed quasars, optical lensed quasars, radio unlensed quasars (with both
flat and steep radio spectra), and radio selected lensed systems. The spectral
indices are calculated from broadband measurements in R and K bands
mostly. While there is not much difference in the spectral indices of lensed
and unlensed optical quasars, the radio samples of lensed and unlensed quasars
differ significantly.}
\end{figure}

\section{Discussion}
Lensed images might also appear red if their light is contaminated by
light from the lensing galaxy, or if the background source is a high
redshift radio galaxy.  These do not appear to be large effects in our
sample: out of 27 lens systems, six are unresolved; in two the
background source is a radio galaxy; and in two the background source
is unidentified. These cases are not the reddest images in the sample,
and removing them from the sample does not change the above
results. Blazars are the reddest flat radio spectrum sources, because
doppler boosted synchrotron radiation contributes to the red part of
their spectra.  The steepest optical-IR spectral index for such
objects (Impey, Lawrence \& Tapia 1991) is $\alpha = -2.6$, a value
which is merely average for lensed radio quasars (Figure 3).  Among
lensed radio sample, only {\it B\,0218+357} is known to be a blazar,
and there is other evidence for extinction in that system (see below).
Starlight from the quasar host galaxy is not important at the
rest-wavelengths (0.6 $\mum$ corresponding to measured K-band) of
these high reshift lensed quasars, because the total light from the
host galaxy is $\sim 2$--$3$ magnitudes fainter than the quasar
(cf. Bahcall, Kirhakos \& Schneider 1996; Mcleod \& Rieke 1995 and
references therein) and not all of the host galaxy is in the strongly
lensed portion of the source plane.  Moreover, the host galaxy light
contributes similarly in radio-loud and radio-quiet quasars (Dunlop et
al. 1993).

\subsection{Estimating the column density of dust}

We could estimate the dust in the lensing galaxy if we knew the intrinsic
spectrum of the source. Lack of knowledge of the reshifts of sources and many
lensing galaxies adds to the uncertainty.  Because the sources are at high
redshifts, measured optical-IR colors correspond to ultraviolet colors in the
rest frame of the sources and optical colors at the redshifts of the lensing
galaxy; hence huge amounts of dust are not needed to produce the observed
reddening. If we assume that the intrinsic spectrum of the quasars is given
by the mean SED (Elvis et al. 1994) and the dust is similar to the Galactic
dust (Draine \& Lee 1984), we estimate $A_{v}= 0.3 - 4$ magnitudes for those
lens systems which have the redshifts of both the lens and the source
measured.

{\it B\,0218+357}  illustrates this point nicely. The ratio of the brightness
of images A/B is much smaller in optical than in radio (Grundahl \& Hjorth
1995). Interferometric observations with the VLA (Menten \& Reid, 1996) show
absorption by ${\rm H_2CO}$ which occurs in dense molecular gas {\it only}
against component A and derive a column density of N(H)$=10^{21}-10^{22}
cm^{-2}$, assuming Galactic abundance of ${\rm H_2CO}$. From CO observations
Wiklind \& Combes (1995) derive a lower limit to the column density of
$N(H_2) > 1.5 \times 10^{21} cm^{-2}$. This, along with HI column density
N(H)$= 10 ^ {18} T_{spin} cm^{-2}$ (Carilli, Rupen \& Yanny 1993), gives
$A_V = 1.9 - 12$ magnitudes for dust and gas similar to Galactic. The
discrepancy between the optical and radio ratios of A and B brightness imply
a differential extinction of 3.7 magnitudes between A and B.  From the
reddening seen in Q0218+357 and assuming the mean SED (Elvis et al. 1994) we
derive a luminosity weighted average extinction in the lens to be 4
magnitudes.

Lawrence et al.  (1995) make a detailed case for {\it MG\,0414+0534} being
obscured by dust, and it is indeed the reddest known radio-selected lens system
followed by {\it 1938+666} (Rhoads, Malhotra \& Turner 1997). Recent
observations of time variation in colors support the dust-in-the-lens
hypothesis for {\it MG\,0414+0534} (Vanderriest et al. 1996). Nadeau et
al. (1991) present evidence for differential reddening of different images of
the quasar in {\it Q\,2237+0305} and derive a reddening law similar to the
Galactic reddening law. Wiklind and Combes (1996) find high (N(H)$= 3 \times
10^{22}$) column density of molecular gas in absorption against one of the
images of {\it PKS\,1830-210} at redshift of 0.89 and Lovell et al. (1996) find
HI in absorption at z=0.19. (The optical counterpart to this lens is not known
and this radio selected lens is not included in our analysis.)

Besides these lenses, there are instances of extinction of background sources
by elliptical galaxies (Stocke et al. 1984, Stickel et al. 1996, McHardy et
al. 1994). To our knowledge there is no systematic study searching for such a
phenomenon at high redshifts, and these cases have been identified because of
some oddity in their behavior at other wavelengths.

\subsection{How dust affects lensing constraints on $\Lambda$}

To see how dust can affect the constraints on $\Lambda$ derived from
statistics of optically selected lens systems, we need to quantify the
differences in color (or spectral indices) between lensed and unlensed
radio selected samples.  This could be done by determining the
transfer function that maps the distribution of spectral indices of
the unlensed sample to that of the lensed sample (top two panels of
figure 3). Ideally this should be done for a large number of lensed
quasars and their parent samples.  The transfer function could then be
applied to the optical quasar sample to see how many reddened and
extincted optical quasars exist and are missing from the surveys.

Such a detailed treatment is not justified at present for three
reasons. First, the transfer function would be noisy because the
number of lensed images is small.  Second, we don't really have the
color distribution of the parent sample of unlensed radio quasars for
all the radio lens surveys (MG, JVAS, CLASS).  And third, if color is
a function of magnitude, we would need to know both the color
distribution of sources down to the intrinsic brightness of the lensed
sources (which is fainter than the survey limits) and the
magnification of each system.

Instead, we use a much simpler procedure of calculating the average
shift in spectral indices between the radio lensed and unlensed
samples. This procedure should give an idea of how much a population
is affected by dust in lenses. The average shift of spectral index,
$\delta \alpha=1.9 \pm 0.27$, translates into different amounts of
dust extinction depending on the redshift of the lens. We estimate an
average extinction of $A_v =1.0 \pm 0.55$ for the radio selected
lensed quasars (where the uncertainty includes the spread in $A_v$ due
to different redshifts of the lensing galaxy).

Next we apply the extinction effects to the calculation of expected
number of lensed quasars in the AAT survey in the optical (Boyle et
al. 1988). Malhotra and Turner (1995) estimated that in the AAT survey
there should be 3.5 lensed systems for a ($\Omega=0.1, \Lambda=0.9$)
cosmology and 0.5 systems for ($\Omega=1, \Lambda=0$) cosmology. This
calculation takes into account the different luminosity functions for
unlensed quasars inferred in different cosmologies and uses reduced
lensing cross-sections from Fukugita \& Turner (1991). An average
extinction of $A_v=1 \pm 0.5$ implies an extinction of $2 \pm 1$
magnitudes of in the U-band, the wavelength at which a $z=0.5$ lensing
galaxy sees the B-band light seen by the observer. The expected number
of lensed systems then drops to 0.25--1.5 systems for a ($\Omega=0.1,
\Lambda=0.9$) cosmology, which is consistent with the AAT survey not
having found a lensed system.

The difference in the inferred reddening between radio-selected and optically
selected lens systems may not be completely attributable to a bias against
finding reddened lens systems in the optical. Radio searches are better able
to detect small-separation lenses, and spiral galaxies contribute about
equally to the lensing optical depth at angular separation of $< 0.5 \arcsec$
(TOG84). So reddening may be more significant for small separation lenses. We
do not, however, see a strong correlation of red images and image separation
in our sample.

\subsection{Confirmation of dust in the lensing galaxies}

The confirmation of dust in lenses could come in various ways, all of which
assume some similarity between dust and gas at high redshifts and in our
Galaxy: (a) Detection of gas in the lensing galaxies suggests dust may be
present, especially if CO is detected, since CO implies reasonable
metallicities.
(b) Detection of differential reddening between images of the same
quasar allows the color difference to be tested for consistency with
known reddening laws, independent of the (unknown) intrinsic quasar
spectrum (cf.\ Nadeau et al.\ 1991).  Because the magnification ratios
and differential dust column densities are imperfectly known, good
photometry of individual lens components in 3 bands is needed for this
test.  Near-IR bands are best for this purpose because the Galactic
reddening law is the least variable at these wavelengths (Whittet
1988). If the relative colors of different images agree with the
Galactic reddening law, one can estimate the differential extinction
between the images. Correcting for this differential extinction will
improve magnification ratio measurements for different images and
hence improve the lens models.
(c) Observation of known, strong spectral features of dust
will perhaps be the most convincing evidence. The $2175\AA $ absorption bump
(cf. Fitzpatrick \& Massa 1986) is a strong feature in the UV and should be
detectable for small column densities of dust (large column densities of dust
would rapidly produce too much extinction in the UV). The silicate absorption
feature at $9.7 \mum$, on the other hand, should be ideal for large column
densities of dust (Roche \& Aitken, 1984) because the continuum extinction at these
wavelengths is small. The absorption features have the advantage that their
detectability depends only on the brightness of the background source (which
is amplified by lensing) and the column density of the dust (which is
exactly what we would like to measure).

\subsection{Conclusions}
The dramatic systematic difference in the optical-IR colors between radio and
optically selected gravitational lens systems and between radio lensed and
unlensed quasars reported here, and the other evidence for dust in lensing
galaxies discussed above, strongly suggest (but do not prove) that optical
searches for lens systems are seriously incomplete due to extinction.  If so,
the limits on $\Lambda$ deduced from lensing statistics from optical surveys
(Maoz \& Rix 1993, Kochanek 1993) are significantly too stringent, and the most
serious objection to invoking non-zero $\Lambda$ models to resolve a variety of
cosmological puzzles (Carroll et al. 1993, Ostriker \& Steinhardt 1994) can be
relaxed.

If the optical searches are biased against images extincted by lenses, and
background radio sources are fewer in number than optical sources, searches in
the near infrared should be most successful in finding lens systems. Stickel et
al. 1996 point out that of a group of 30 radio sources picked out for being
unusual in their steep optical index by Rieke et al. (1979, 1982) and Smith \&
Spinrad (1980) two (B 0218+357 and MG 0414+0534) were later identified as being
gravitationally lensed. Usual fraction of lenses in an optical or radio surveys
is about 1 or 2 systems per 1000 quasars! The reddened systems can be used to
study the interstellar medium and dust at high redshifts, as is being done now
(Wiklind \& Combes, 1995, 1996, Carilli et al. 1993, Menten \& Reid, 1996,
Nadeau et al. 1991). Gravitational lensing thus has the potential to contribute
to our knowledge of interstellar matter and dust in the high redshift galaxies,
as well as the gravitational potential of galaxies and clusters and the
geometry of the universe.

\section*{Acknowledgements} 
We thank Carol Lonsdale and Roc Cutri for providing electronic tables of
optical infrared colors of PG quasars; Chris Kochanek and Joseph Lehar for
providing finding charts in advance of publication; Charles Lawrence for
discussions and Joachim Wambsganss and Chris Kochanek for comments on an
earlier draft of the paper.  We also thank the staff of Apache Point
Observatory for their help during these observations. The Apache Point
Observatory is privately owned and operated by Astrophysical Research
Consortium, consisting of the University of Chicago, Institute for Advanced
Study, Johns Hopkins University, New Mexico State University, Princeton
University, University of Washington and Washington State University This
work was partly supported by the NSF grant AST94-19400 and GER9354937.

\end{document}